\documentclass[pre, aps, twocolumn, reprint, amsmath,amssymb, showpacs,
superscriptaddress] {revtex4-1}

\usepackage{amsmath}    % need for subequations
\usepackage{graphicx}   % need for figures
\usepackage[colorlinks=true, citecolor=blue, linkcolor=blue]{hyperref}   % use for hypertext links, including those to external documents and URLs

\usepackage{dcolumn}% Align table columns on decimal point
\usepackage{bm}% bold math
\begin{document}

\title{Physical interpretation of the partition function for colloidal clusters}
\date{\today}
\author{Ellen D. Klein}
\affiliation{Department of Physics, Harvard University, 17 Oxford
  Street, Cambridge, Massachusetts 02138}
\author{Rebecca W. Perry}
\affiliation{Harvard John A. Paulson School of Engineering and Applied Sciences,
  Harvard University, 29 Oxford Street, Cambridge, Massachusetts 02138}
\author{Vinothan N. Manoharan}%
 \homepage{http://manoharan.seas.harvard.edu}
 \email{vnm@seas.harvard.edu}
\affiliation{Harvard John A. Paulson School of Engineering and Applied Sciences,
  Harvard University, 29 Oxford Street, Cambridge, Massachusetts 02138}
\affiliation{Department of Physics, Harvard University, 17 Oxford
  Street, Cambridge, Massachusetts 02138}

\begin{abstract}\label{sec:abstract}
  Colloidal clusters consist of small numbers of colloidal particles
  bound by weak, short-range attractions. The equilibrium probability of
  observing a cluster in a particular geometry is well-described by a
  statistical mechanical model originally developed for molecules. To
  explain why this model fits experimental data so well, we derive the
  partition function classically, with no quantum mechanical
  considerations. Then, by comparing and contrasting the derivation in
  particle coordinates with that in center-of-mass coordinates, we
  physically interpret the terms in the center-of-mass formulation,
  which is equivalent to the high-temperature partition function for
  molecules. We discuss, from a purely classical perspective, how and
  why cluster characteristics such as the symmetry number, moments of
  inertia, and vibrational frequencies affect the equilibrium
  probabilities.
\end{abstract}

\maketitle

\section{Introduction\label{sec:introduction}}
A colloidal cluster consists of a small number of colloidal particles,
often spherical, that are held together by short-range attractions.
Experimentally, such systems can be made by isolating small numbers of
colloidal microspheres in two~\cite{perry_two-dimensional_2015,
  perry_segregation_2016} or three dimensions~\cite{meng10, perry12} in
the presence of micelles or small particles, which induce a depletion
attraction \cite{asakura54, vrij76, lekkerkerker11} between the
microspheres. When the attractive interactions are weak, the particles
can rearrange into different configurations on experimental time scales.
Studies of these configurations yield insights into nucleation
barriers~\cite{hoy12, perry12}, the glass
transition~\cite{yunker_relationship_2013, hoy15}, and the emergence of
a phase transition as the size of a system increases~\cite{perry12,
  calvo12}.

Over the past few years, the minimal-energy configurations of small
colloidal clusters have been studied extensively in experiment, theory,
and simulation~\cite{malins_geometric_2009, arkus09, meng10, wales10,
  hoy_minimal_2010, arkus_deriving_2011, hoy12, perry12, calvo12,
  morgan14, perry_two-dimensional_2015, hoy15,
  holmes-cerfon_enumerating_2016, kallus_free_2017,
  holmes-cerfon_sticky-sphere_2017}. We and others~\cite{meng10,
  wales10, morgan14, perry_two-dimensional_2015} have found that a
statistical mechanical model originally developed for molecules can
accurately predict the equilibrium occurrence frequencies of the
minimal-energy structures. In some ways, the agreement makes sense: the
particles have well-defined interactions, are small enough to display
Brownian motion, and can reach thermal equilibrium on experimental
timescales. There is no reason statistical mechanics \emph{shouldn't}
describe their properties.

But it is perplexing that a molecular model usually derived from quantum
mechanical arguments can so accurately predict the properties of a
purely classical system. Typical colloidal particles are around a
micrometer in diameter, or 10,000 times the diameter of a hydrogen atom.
Unlike the atoms that make up molecules, the particles that make up
clusters are in principle distinguishable, since each particle contains
a different number of molecules or has a different size. Even the
rotations of spherical colloidal particles are---again, in
principle---observable; the particles might have a small optical
anisotropy or a slight eccentricity that can be used to measure
orientation. None of these features are taken into account in the
molecular model. Why and how does it describe classical systems?

To answer this question, we derive the partition function for colloidal
clusters, starting from classical statistical mechanics and leaving out
all quantum mechanical considerations. Our goal is to clarify the
underlying physics; more rigorous and general derivations can be found
in the work of Holmes-Cerfon and
coworkers~\cite{holmes-cerfon_enumerating_2016,
  holmes-cerfon_sticky-sphere_2017, kallus_free_2017}. We use our
derivations to explain how properties such as the symmetry number,
moments of inertia, and vibrational frequencies affect the equilibrium
probability of observing a particular cluster structure. The roles of
these properties are often interpreted in terms of quantum mechanics or
dynamics, but, as we shall show, their effects can be understood in
terms of classical physics and geometry.

%%%%%%%%%%%%%%%%%%%%%%%%%%%%%%%%%%%%
\subsection{Background}
\label{sec:background}

To motivate our work, we first describe the equilibrium between two
cluster structures with $N=6$ spherical particles. The equilibrium ratio
of the two structures was explored in simulation by Malins and coworkers
\cite{malins_geometric_2009} and in experiment by Meng and coworkers
\cite{meng10} and Perry and coworkers~\cite{perry12}. The experiments
used micrometer-scale spherical particles that were held together by
short-range, attractive depletion interactions.

\begin{figure}
  \begin{centering}
  \includegraphics{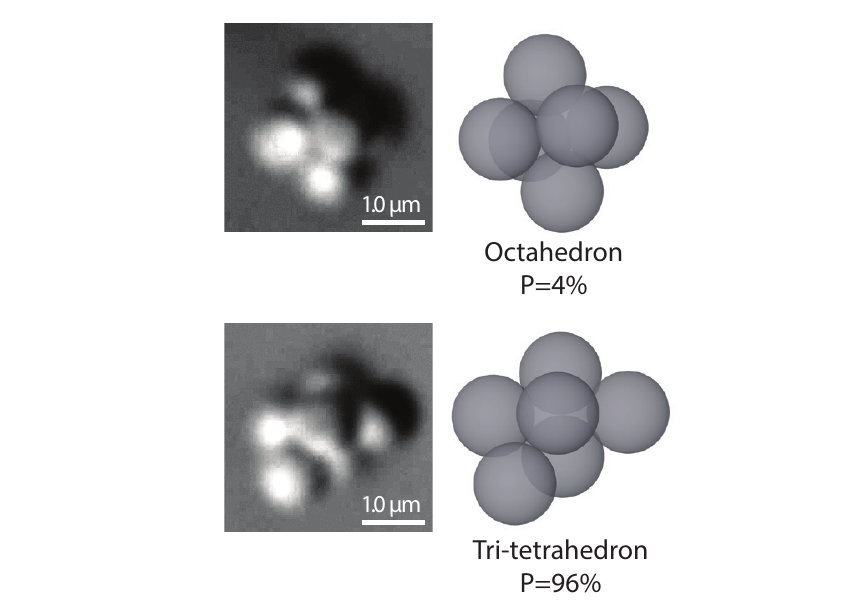}
  \caption{Equilibrium between octahedral and tri-tetrahedral
    structures. Meng and collaborators experimentally observed the
    tri-tetrahedron (bottom) 24 times as often as the octahedron
    (top)~\cite{meng10}. This difference is due primarily to the
    tri-tetrahedron's higher rotational entropy. \label{fig:Octa-Poly}}
  \end{centering}
\end{figure}

For the six-particle system, there are two structures that minimize the
total potential energy: the octahedron and tri-tetrahedron
(Fig.~\ref{fig:Octa-Poly}). Both have the same number of interacting
pairs of spheres (``bonds'') and hence the same potential energy, but
the tri-tetrahedron occurs 24 times more often in an equilibrium
ensemble.

To understand why the tri-tetrahedron occurs so much more often, Meng
and colleagues used a statistical mechanical model originally developed
for molecules, in which the total partition function $Q'$ is written as
the product of partition functions for collective translations,
rotations, and vibrations: $Q' =
Q'_\textnormal{trans}Q'_\textnormal{rot}Q'_\textnormal{vib}$. Each term
represents a different entropic contribution to the free energy. Because
the partition function is proportional to the probability of observation
in equilibrium, the ratio of the partition functions for the
tri-tetrahedron and octahedron should be 24:1.

Meng and coworkers found that the largest contribution to the factor of
24 comes from a factor called the symmetry number, which accounts for
all the permutations of particles that lead to the same structure. The
number of ways in which six particles can form an octahedron, which has
multiple axes of fourfold, threefold, and twofold symmetry, is much
smaller than the number of ways in which six particles can form the
tri-tetrahedron, which has only one axis of twofold symmetry. Thus, the
octahedron has a much larger symmetry number than the tri-tetrahedron.
In equilibrium, the tri-tetrahedron is therefore favored by a factor of
12, corresponding to the ratio of symmetry numbers. We discuss the
origin of the symmetry number and its physical interpretation in more
detail in Sections~\ref{sec:stat-mech-model} and~\ref{sec:discussion}.

The remaining factor of two comes from a term in the rotational
partition function that is proportional to the product of the moments of
inertia, which differs between the two structures, and the vibrational
partition function, which can be calculated using a harmonic
approximation for the potential.

The model can be generalized to two-dimensional systems \cite{morgan14,
  perry_two-dimensional_2015} and to clusters with $N>6$ particles,
where the number of minimal-energy structures increases rapidly with $N$
\cite{meng10, perry12, kallus_free_2017}. One interesting result from
these studies is the dominance of symmetry effects when $N$ is small:
Meng and coworkers found that when $N<9$, the clusters always favor
asymmetric configurations in equilibrium.

\subsection{Overview}
\label{sec:overview}

In what follows, we explain why the partition function can be written in
the form above, and how the factors that appear in the rotational and
vibrational parts affect the equilibrium probabilities. To do this, we
first introduce the elements and assumptions of our model in
Section~\ref{sec:framework} and then derive the partition function in
two different coordinate systems: particle coordinates
(Section~\ref{sec:Q_in_particle_coords}) and center-of-mass coordinates
(Section~\ref{sec:Q_in_COM}). The formulation in particle coordinates
does not lend itself to analytical calculations, whereas that in
center-of-mass coordinates can be used to explicitly calculate the
observation probabilities. However, the derivation in particle
coordinates is more general, and we use it to gain physical insights
into the terms in the center-of-mass formulation. In the discussion
(Section~\ref{sec:discussion}) we equate the two versions to explain
the origin and roles of the symmetry number and the dynamical quantities
that appear in the center-of-mass formulation---the moments of inertia
and the vibrational frequencies.

%%%%%%%%%%%%%%%%%%%%%%%%%%%%%%%%
\section{The statistical mechanical model}
\label{sec:stat-mech-model}

\subsection{Framework}
\label{sec:framework}

\begin{figure}
	\begin{centering}
    \includegraphics{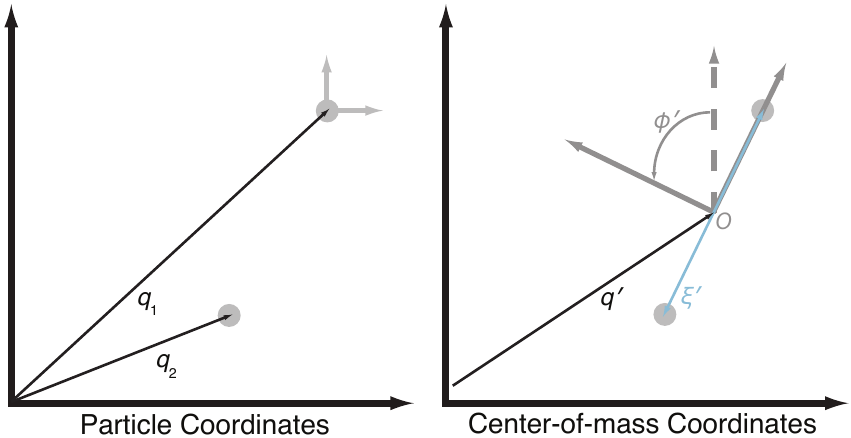}
    \caption{In particle coordinates (left) the positions of every
      particle ($\mathbf{q}_i$) are measured from the origin of a lab
      frame. In center-of-mass coordinates (right), we define a rotating
      frame (dark gray) with an origin $O$ at the cluster's center of
      mass; the position of $O$ relative to the lab frame is given by
      $\mathbf{q'}$. The rotating frame is chosen to lie along the
      cluster's principal axes. The standard Euler angles (the first of
      which is $\phi'$) describe its rotation relative to the lab frame.
      Within the rotating frame, the coordinates of the vibrational
      modes are denoted by $\xi'$. Particles are also free to rotate
      about their own centers of mass. Describing these rotations
      requires another rotating coordinate system located at the center
      of mass of each particle (light gray axes on left).
      \label{fig:CoordSystems}}
	\end{centering}
\end{figure}

We seek a model for the experimental observable $P_{s_k}$, the
probability of observing a particular structure $s_k$ in an equilibrium
ensemble. For example, in the $N=6$ case discussed above, there are two
structures: $s_1=\textnormal{octahedron}$ and
$s_2=\textnormal{tri-tetrahedron}$. In equilibrium, $P_{s_k}$ is
proportional to $Q_{s_k}$, the partition function of $s_k$: 
\begin{equation}
	P_{s_k} = \frac{Q_{s_k}}{\sum_{l}{Q_{s_l}}},
  \label{eq:cluster_probability}
\end{equation}
where the summation ranges over all structures $s_l$ in the ensemble. In
experiments, one usually counts only clusters that represent minima in
the energy---that is, those with at least $3N-6$ bonds---as part of the
ensemble. States with fewer bonds are ignored. We calculate the
partition function in the two coordinate systems illustrated in
Fig.~\ref{fig:CoordSystems}: particle coordinates and center-of-mass
coordinates.

\subsubsection{Interactions}

We assume that our system is at constant temperature and that the
interactions between particles are pairwise additive and spherically
symmetric. We also assume that the potential is short-ranged. These are
good approximations for the experimental systems discussed above:
micrometer-scale electrostatically-stabilized particles subject to
depletion interactions in water at moderate to high salt concentrations.
The repulsions are short-ranged because the salt screens electrostatic
interactions. The depletion attraction is short-ranged because the
particles that cause it are typically much smaller than the diameter $d$
of the colloidal particles.

\subsubsection{Degrees of freedom}

We define the phase space of our system by the positional degrees of
freedom and their conjugate momenta. We implicitly account for the
degrees of freedom of the solvent molecules by using a potential of mean
force to describe the interactions between the particles. This potential
is a thermal average over all the configurations of solvent
molecules~\cite{poon_introduction_2006}. Therefore the phase space is
determined by the degrees of freedom of the particles alone.

\begin{figure}
	\begin{centering}
    \includegraphics{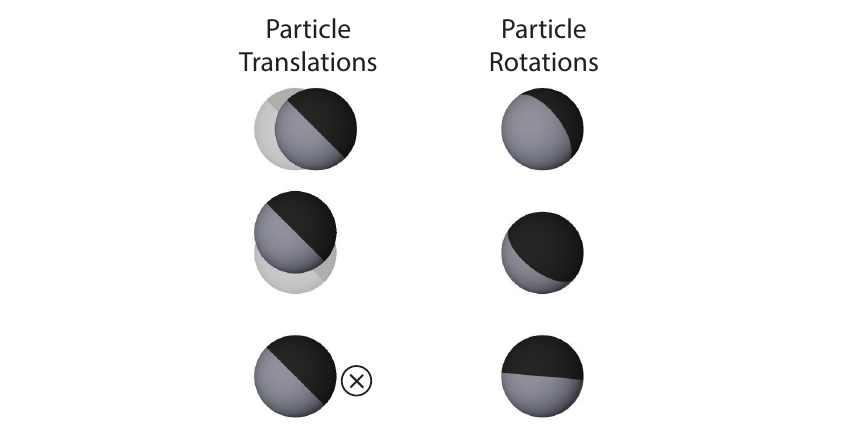}
    \caption{An individual colloidal particle has six positional degrees
      of freedom: three translational and three rotational.  The three translational
      degrees of freedom have three conjugate linear momenta, and the three
      rotational degrees of freedom have three conjugate angular momenta.
      The rotational degrees of freedom might be observed by watching
      small defects on the surfaces of the particles or by dyeing one
      hemisphere of each particle, as illustrated here.
      \label{fig:ParticleDOF}}
	\end{centering}
\end{figure}

To illustrate how the degrees of freedom differ in the two coordinate
systems, we consider a dimer. In the particle coordinate system, the
dimer has 12 positional degrees of freedom: each particle can translate
in each of the three dimensions, and each can rotate about three
independent axes centered on its center of mass. The rotational motions
can in principle be observed by tracking small defects on the surfaces
of the particles or by dyeing part of each particle, as shown in
Fig.~\ref{fig:ParticleDOF}. Interactions such as depletion change the
distribution of values for each degree of freedom relative to a gas, but
they do not change the number or type of degrees of freedom.

\begin{figure}
\centering
\includegraphics{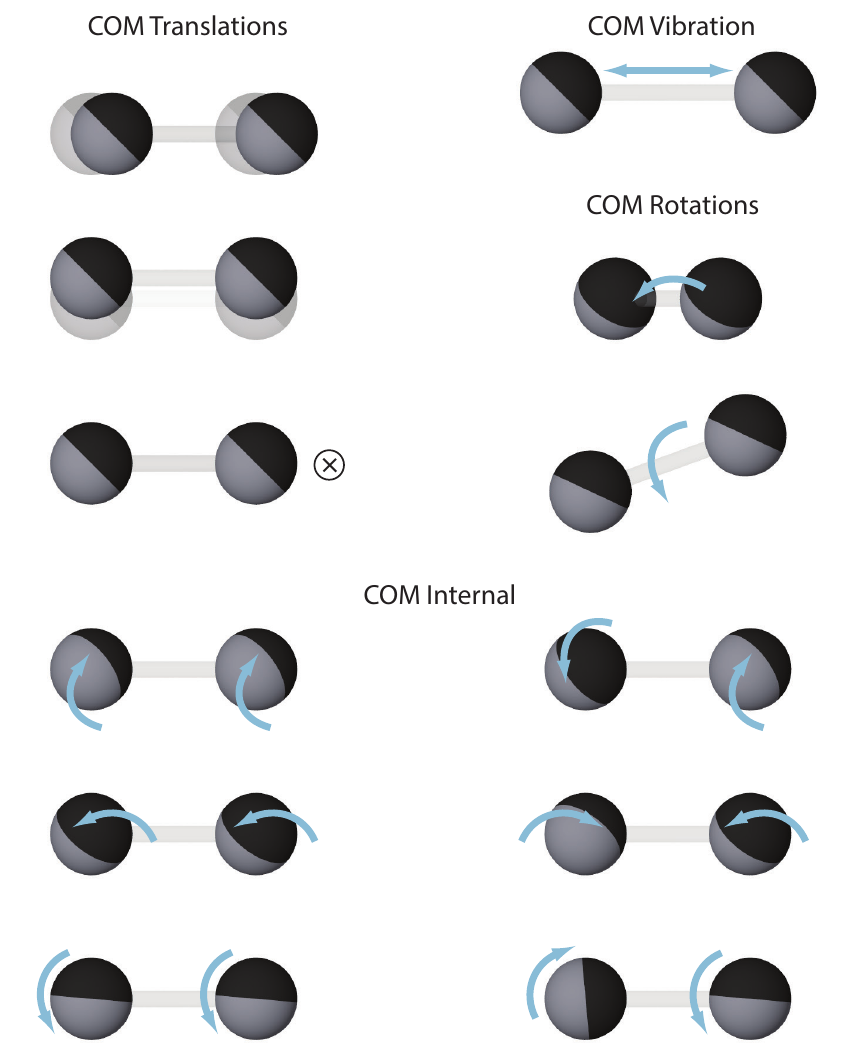}
\caption{\label{fig:dimer-DOF}In center-of-mass coordinates, a colloidal
  dimer has 12 degrees of freedom, three corresponding to full-body
  translations, two to full body rotations, one to vibration, and six
  others to internal modes. The internal degrees of freedom arise from
  the rotations of individual colloidal particles.}
\end{figure}

In the center-of-mass coordinate system, the dimer also has 12 degrees
of freedom (Fig.~\ref{fig:dimer-DOF}). Three correspond to translations
of the center of mass, two to rotations of the cluster about its center
of mass, one to vibrations of the bond, and the remaining six to
internal modes. An internal mode is one where particles rotate about
their own centers of mass, either in the same direction or in the
opposite direction as their partners. The top left internal mode in
Fig.~\ref{fig:dimer-DOF} (bottom) is equivalent to rotations of the
entire dimer about its axis.

Importantly, none of these modes can be ``frozen out,'' as might happen
in a molecular system. In a diatomic molecule such as $N_2$, the excited
vibrational states are not accessible at room temperature, because the
energy levels are much larger than the thermal energy. In the classical
dimer, all 12 modes can be excited, and we account for all of them in
our derivation. We do, however, neglect modes associated with vibrations
of the molecules \emph{inside} the particles.

\subsubsection{Distinguishability}

Whereas in a molecule like $N_2$, the two nitrogen atoms are
fundamentally indistinguishable (if they are the same isotope), in a
colloidal system the particles are distinguishable, as discussed above.
However, we can choose \emph{not} to distinguish the particles from one
another. This is a common---if not universal---tactic used in the
analysis of experiments on colloidal
self-assembly~\cite{cates_celebrating_2015}. The term
\emph{undistinguished}, coined by Sethna~\cite{Sethna_2006}, describes
the particles in this situation. We assume undistinguished particles
throughout.

%%%%%%%%%%%%%%%%%%%%%%%%%%%%%
\subsection{Partition function in particle coordinates}
\label{sec:Q_in_particle_coords}

In particle coordinates, each particle has six positional degrees of
freedom (Fig.~\ref{fig:ParticleDOF})---three translational and three
rotational---and six associated momenta---three linear and three
angular. Thus, a cluster of $N$ particles has $6N$ positional degrees of
freedom and $6N$ associated momenta.

The translational degrees of freedom for the $i$th particle $(q_{ix},
q_{iy}, q_{iz})$ are measured as displacements from the origin
(Fig.~\ref{fig:CoordSystems}, left), which is fixed in the lab frame.
The set of all translational degrees of freedom is $\mathbb{Q} =
(q_{1x}, q_{1y}, q_{1z}, \ldots, q_{Nz})$. The linear momenta
corresponding to the translational degrees of freedom for the $i$th
particle are $\mathbf{p}_i = (p_{ix}, p_{iy}, p_{iz})$, and the set of
all linear momenta is $\mathbb{P}$.

Each particle can also rotate about its own center of mass. We describe
these rotations using a rotating frame with its origin at the particle's
center of mass (light gray axes in Fig.~\ref{fig:CoordSystems}, left).
Each particle $i$ can rotate through Euler angles $(\phi_{i},
\theta_{i}, \psi_{i})$ relative to the lab frame. The set of all such
angles is $\bm{\Phi}=(\phi_1, \theta_1, \psi_1, \ldots, \phi_N,
\theta_N, \psi_N)$. The angular kinetic energy of the individual
particles depends on the set of all momenta $\mathbb{L}$ conjugate to
the Euler angles.

With the definitions above, we can express the Hamiltonian $\mathcal{H}$
for a system of $N$ particles as
\begin{equation}\label{Particle-Hamiltonian}
  \begin{split}
    \mathcal{H} &= 
    U(\mathbb{Q}) + K(\mathbb{P}) + U(\bm{\Phi})
    + K(\mathbb{L}),
  \end{split}
\end{equation}
where $U$ is potential energy---again, a potential of mean force---and
$K$ is kinetic energy. The canonical partition function $Q$ is then
\begin{equation}\label{particle_prop_partitionfunction}
  Q \propto \int {e^{-\beta \mathcal{H} } }d\mathbb{Q} \, d\mathbb{P} \, d\bm{\Phi} \, d\mathbb{L},
\end{equation}
where $\beta=1/(k_B T)$, $k_B$ being Boltzmann's constant and $T$ the
temperature of our system. We use a proportionality symbol because we
have yet to determine the bounds and the prefactors.

The bounds on the integral must be consistent with our definition of the
structure $s$. If we were to integrate over all of phase space, then the
partition function would include all possible structures. Instead, we
integrate only over those parts of phase space in which the particles
are arranged in a particular structure $s$.

One way to define the structure is through an adjacency matrix
$\mathbf{A}_s$~\cite{arkus09, arkus_deriving_2011}, a symmetric,
$N\times N$ matrix. An element $A_{ij}$ is equal to 1 if particle $i$ is
bound to particle $j$, and 0 otherwise. To determine whether two
particles are bound, we must first set a cutoff distance $\ell$. For
instance, $\ell$ might be the maximum range of the depletion force. For
a short-range interaction, $(\ell-d)/d \ll 1$. This definition requires
assigning a unique label to each particle in our structure.

We would like the partition function for a structure $s$ to integrate
over all fluctuations of that structure, because experiments do not
distinguish structures by their center-of-mass positions, orientations,
or distances between particles (as long as the center-to-center distance
between particles is less than $\ell$). The adjacency matrix
$\mathbf{A}_s$ is a convenient way to delineate the bounds on phase
space because it describes the structure irrespective of such
fluctuations. Therefore, if we set the bounds on the integral in
Eq.~(\ref{particle_prop_partitionfunction}) to include the region of
phase space in which the adjacency matrix is $\mathbf{A}_s$, the
partition function will include contributions from the rotations of the
individual particles, translations of the entire cluster, rotations of
the entire cluster, and fluctuations in interparticle distances.

\begin{figure}
	\begin{centering}
    \includegraphics{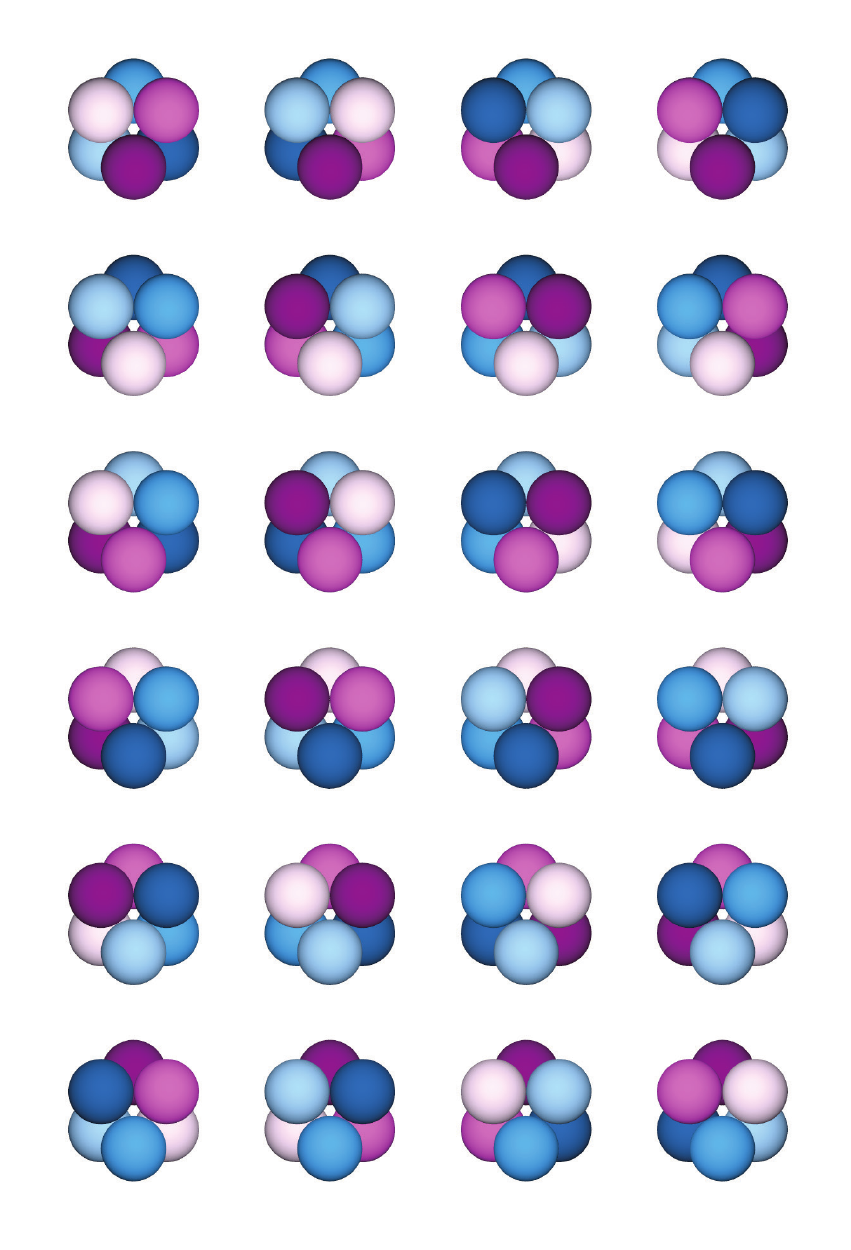}
	\caption{24 of the 720 colorings, or label permutations, of the
    octahedron. All of these colorings are equivalent through
    rotations.
    \label{fig:Octahedron_Colorings}}
	\end{centering}
\end{figure}

However, some of these \emph{fluctuations} correspond to different
\emph{representations} of the same structure. The overlap arises because
the adjacency matrix is not a unique representation of a structure.
There are $N!$ different adjacency matrices that correspond to the same
structure because there are $N!$ permutations of particle labels. Some
of these permutations are identical to other permutations plus full-body
rotations, as illustrated in Fig.~\ref{fig:Octahedron_Colorings}.
Therefore, for any given representation of the structure (any particular
adjacency matrix), we must divide the partition function by a factor
that accounts for how many orientations are shared with a different
representation. That factor is the symmetry number $\sigma_s$. It is
equal to 24 for the octahedron, as shown in
Fig.~\ref{fig:Octahedron_Colorings}.

The symmetry number $\sigma_s$ accounts for all the ways in which
permutations plus rotations yield an identical cluster. We discuss
$\sigma_s$ in more detail in section \ref{sec:SymmetryNumber}. We note
here that because $\sigma_s$ appears even in our classical derivation,
it cannot arise from any quantum mechanical considerations. As noted by
Gilson and Irikura~\cite{gilson_symmetry_2010}, $\sigma_s$ is a
mathematical artifact arising from how we define the region of phase
space that we integrate over. Indeed, it can be calculated from the size
of the automorphism group of the adjacency
matrix~\cite{arkus_deriving_2011}.

A further complication is that a given adjacency matrix can correspond
to chiral enantiomers~\cite{arkus_deriving_2011} or two or more
geometrically-distinct clusters~\cite{holmes-cerfon_enumerating_2016}.
Therefore, if we were to integrate over the regions of phase space
corresponding to one such matrix, we would include contributions from
structures that an experimentalist might treat as different. However, we
will not use the partition function in particle coordinates to evaluate
the equilibrium probabilities; we use this version only to gain insight
into the partition function in center-of-mass coordinates, which is much
more tractable. With this aim in mind, we restrict our discussion to
only those cases in which the adjacency matrix defines a single
structure.

Finally, we must include a prefactor of $1/h^{6N}$ for dimensional
consistency, where $h$ is a placeholder for any quantity with dimensions
of momentum times length. The exponent of $6N$ arises because there is
one factor of $h$ for each conjugate pair of position and momentum in
phase space. We do not claim---nor do we need to claim---that $h$ is
Planck's constant, since the quantity $h$ must cancel in the statistical
mechanical calculation of \emph{any} classical observable. It can appear
only if a degree of freedom is frozen out, in which case the calculation
is no longer classical.

The resulting partition function for a structure $s$ is
\begin{equation}
  Q_s = \frac{1}{\sigma_s h^{6N}} \int_{\mathbf{A}_s}{e^{-\beta
      \mathcal{H} } }d\mathbb{Q} \, d\mathbb{P} \, d\bm{\Phi} \,
  d\mathbb{L},
\end{equation}
where the subscript $\mathbf{A}_s$ reminds us that the integral is over
the region of phase space corresponding to just one labeling.

The separability of the Hamiltonian in Eq.~\ref{Particle-Hamiltonian}
allows us to factor the partition function into configurational and
momentum components:
\begin{equation}\label{ParticleCoords-GeneralPartitionFunction}
  \begin{split}
    Q_s = \frac{1}{\sigma_s h^{6N}} &\int_{\mathbf{A}_s}{e^{-\beta \mathcal{H} } }d\mathbb{Q} \, d\mathbb{P} \, d\bm{\Phi} \, d\mathbb{L}\\
    = \frac{1}{\sigma_s h^{6N}} & \int_{\mathbf{A}_s}{e^{- \beta U(\mathbb{Q}) } }d\mathbb{Q}
    \int{e^{- \beta K(\mathbb{P}) } }d\mathbb{P}\\
    &\times\int{e^{- \beta U(\bm{\Phi}) } }d\bm{\Phi}
    \int{e^{- \beta K(\mathbb{L}) } }d\mathbb{L}\\
    = \frac{1}{\sigma_s h^{6N}} &Q_{s, \textnormal{trans}} (\mathbb{Q}, \mathbb{P}) \, Q_{s, \textnormal{rot}}(\bm{\Phi}, \mathbb{L}),
  \end{split}
\end{equation}
where the last line defines the translational ($Q_{s,
  \textnormal{trans}}$) and rotational ($Q_{s, \textnormal{rot}}$)
components of the partition function in particle coordinates. Note that
the terms ``translational'' and ``rotational'' refer to the degrees of
freedom of individual particles, not of the center of mass of the entire
cluster. Note also that this decomposition holds for all classical
systems, because the positions and the momenta always decouple in the
classical Hamiltonian. The adjacency matrix determines the bounds only
on the integral over $\mathbb{Q}$. The bounds on the linear-momentum
integral extends from $-\infty$ to $+\infty$, and the bounds on the
integrals defining $Q_{s, \textnormal{rot}}$ extend over all Euler
angles and associated momenta.

%%%%%%%%%%%%%%%%%%%%%%%%
\subsubsection{Translations and linear momenta}

We first examine the part of the partition function corresponding to
translations of individual particles. From
Eq.~(\ref{ParticleCoords-GeneralPartitionFunction}),
\begin{equation}\label{Particle_TranslationalPartitionFunction}
	Q_{s, \textnormal{trans}} = \int_{\mathbf{A}_s}{e^{- \beta U(\mathbb{Q}) } }d\mathbb{Q}
  \int {e^{- \beta K(\mathbb{P}) } }d\mathbb{P}. 
\end{equation}

To understand how the structure affects the first integral, we assign
effective volumes to each particle in the cluster. We can think of the
first particle as free to wander the entire volume $V$ of the container.
The second particle, which is bound to the first, is then constrained to
a spherical shell around the first particle with inner radius $d$ and
thickness $\ell-d$. The effective volume corresponding to this particle
depends on the interaction potential, which weights the different
regions of the shell. A third particle would be similarly constrained to
an effective volume defined by the other particles, and so on.

We can therefore write the configurational partition function $Z_s$ as a
product of volumes~\cite{herschbach59}:
\begin{equation}\label{Particle-ConfigurationalPartitionFunction}
	Z_s = \int_{\mathbf{A}_s}{e^{-\beta U(\mathbb{Q})}}d\mathbb{Q} =  V\prod_{i=2}^{N}V_{s,i},
\end{equation}
where $V_{s,i}$ is the effective volume that the $i$th particle is
allowed to explore in our structure $s$. Here we have assumed that the
volume of the container $V$ is much larger than the volume of a single
particle. For certain structures, these effective volumes can be
calculated explicitly by transforming the integral in
Eq.~(\ref{Particle-ConfigurationalPartitionFunction}) to internal
coordinates~\cite{herschbach59, boresch96}.

We then integrate over the momenta. The kinetic energy is that for a
non-relativistic classical system:
\begin{equation*}\label{Particle-TranslationalKineticEnergy}
	\begin{split}
    K(\mathbb{P}) =\sum_{i=1}^{N} {\frac{1}{2 m_i} (p_{i_x}^2+p_{i_y}^2+p_{i_z}^2)}.
    \end{split}
\end{equation*}
Thus, the translational part of the partition function in particle
coordinates is
\begin{equation}\label{Particle-TranslationalPartitionFunction}
	\begin{split}
		Q_{s, \textnormal{trans}} & = \int_{\mathbf{A}_s}{e^{-\beta U(\mathbb{Q})}}d\mathbb{Q} \int{e^{-\beta K(\mathbb{P}) } }d\mathbb{P} \\
    	& = Z_s \int{\exp{\left(-\beta\sum_{i=1}^{N} \frac{p_{i_x}^2+p_{i_y}^2+p_{i_z}^2}{2 m_i}  \right)}}d\mathbb{P} \\
        & = Z_s \prod_{i=1}^{N} \left({\frac{2 \pi m_i}{\beta}} \right)^{3/2},
	\end{split}
\end{equation}
where the last line follows from evaluating the Gaussian integrals for
each momentum component. We note that because our (non-gravitational)
potential does not depend on the particle masses, $Z_s$ also does not
depend on the masses.

%%%%%%%%%%%%%%%%%%%%%%%%
\subsubsection{Rotations and angular momenta}

Finally, we turn to the rotational component of the partition function.
From Eq.~(\ref{ParticleCoords-GeneralPartitionFunction}), this component
is
\begin{equation*}
	Q_\textnormal{s, rot}=\int{e^{- \beta U(\bm{\Phi}) } }d\bm{\Phi}
    \int{e^{- \beta K(\mathbb{L}) } }d\mathbb{L}.
\end{equation*}
It accounts for the rotation of particles about their own centers of
mass.

We assume that the rotational potential energy depends neither on the
orientations of the particles nor on their positions in the cluster.
Hence we can say $U(\bm{\Phi})=0$. This is a good approximation for
spherical colloidal particles subject to depletion interactions, which
are isotropic and short-ranged. The rotational kinetic energy of the
cluster is the sum of that of the individual particles, and so the
rotational component of the partition function is the product of the
rotational components of each particle.

Therefore, $Q_\textnormal{s, rot}$ is a constant that depends on the
number of particles $N$ and their moments of inertia, but not the
structure $s$. We therefore drop the subscript $s$ on the rotational
part and let $Q_\textnormal{s, rot} = Q_\textnormal{rot}$. Because
$Q_\textnormal{rot}$ cancels when we calculate the probability of
observing a structure $s$ from Eq.~(\ref{eq:cluster_probability}), we
need not calculate it explicitly.

%%%%%%%%%%%%%%%%%%%%%%%%
\subsubsection{Complete partition function in particle coordinates}

The complete partition function for a structure $s$ in particle
coordinates is
\begin{equation}\label{Particle-CompletePartitionFunction}
	\begin{split}
		Q_s & = \frac{1}{\sigma_s h^{6N}} Q_{s, \textnormal{trans}} Q_{\textnormal{rot}} \\
    & = \frac{ Q_{\textnormal{rot}} }{\sigma_s h^{6N}} Z_{s} \prod_{i=1}^{N} \left({\frac{2 \pi m_i}{\beta}} \right)^{3/2} \\
    & = \frac{ Q_{\textnormal{rot}}}{\sigma_s h^{6N}} \left( \frac{2 \pi}{\beta} \right)^{3N/2}  \left( V\prod_{i=2}^N V_i  \right) \prod_{i=1}^{N} m_i^{3/2}.
  \end{split}
\end{equation}
The version for quasi-two-dimensional systems is given in the Appendix.

The probability of observing structure $s_1$ relative to that of $s_2$,
where both structures have the same $N$ particles, is
\begin{equation}
	\frac{P_{s_1}}{P_{s_2}}=\frac{Q_{s_1}}{Q_{s_2}} = \frac{\sigma_{s_2}}{\sigma_{s_1}}\frac{Z_{s_1}}{Z_{s_2}}
  \label{eq:cluster_probability_particle_coords}
\end{equation}
where $Q_\textnormal{rot}$ has canceled because it does not depend on
$s$. The ratio of the probabilities is therefore inversely proportional
to the ratio of symmetry numbers and directly proportional to the ratio
of $Z_s$, which is the product of effective volumes.

Equation~(\ref{eq:cluster_probability_particle_coords}) has a
straightforward physical interpretation. Structures with greater
flexibility or range of internal motion are favored in equilibrium
because they have larger effective volumes or, equivalently, larger
$Z_s$. As we discuss below, $Z_s$ is related to vibrations and rotations
in center-of-mass coordinates.
Equation~(\ref{eq:cluster_probability_particle_coords}) also shows that
structures with low symmetry are favored over those with high symmetry.
We discuss this effect in Section~\ref{sec:SymmetryNumber}.

Lastly, we note that the masses of the individual particles, even if
different, have no effect on the ratio of equilibrium probabilities, as
long as the total masses of the clusters are the same. The masses cancel
from the probability ratio under the assumptions we have made.

%%%%%%%%%%%%%%%%%%%%%%%%
\subsection{Partition function in center-of-mass coordinates}
\label{sec:Q_in_COM}

Calculation of the equilibrium probabilities is simpler in the
center-of-mass coordinate system because we can make the
rigid-rotor-harmonic-oscillator approximation. Below, we explain and
justify this approximation and then derive the partition function. We
use a prime ($'$) symbol to denote all quantities defined in
center-of-mass coordinates.

The rigid-rotor-harmonic-oscillator approximation allows us to separate
the Hamiltonian into terms that describe the translation of the center
of mass, rotations about the center of mass, and vibrations of particles
about their lowest-energy (equilibrium) positions~\cite{Wilson_1955}.
For this approximation to hold, the amplitude of the vibrational motion
must be small compared to the equilibrium distance between particle
centers. In that case, we can treat the rotations of the cluster using
rigid-body mechanics and the vibrations using a normal-mode framework.

We justify this approximation on three grounds. First, we expect the
vibrational motion to be small because the interactions are short-ranged
for a typical colloidal system. Second, Meng and coworkers~\cite{meng10,
  meng_elastic_2014} showed that a harmonic potential is a reasonable
approximation for the combination of a depletion interaction and
electrostatic repulsion. Third, and most importantly, Meng and
colleagues showed that the predictions of a statistical mechanical model
based on the rigid-rotor-harmonic-oscillator approximation gave
excellent agreement with experiment.

In applying this approximation, we must limit our analysis to those
clusters that rotate as rigid bodies. We therefore exclude ``singular''
clusters, which are minima of the potential energy containing at least
$3N-6$ pair interactions but which contain zero-frequency vibrational
modes. Kallus and Holmes-Cerfon have shown how to calculate the free
energy for these clusters~\cite{kallus_free_2017}. We also exclude
hyperstatic clusters---those with more than $3N-6$ pair interactions. We
restrict the derivation to non-singular isostatic clusters ($3N-6$ pair
interactions with no zero-frequency modes) because our primary goal is
to give physical insight into the form of the partition function.

We can describe a non-singular, non-hyperstatic cluster of $N$ particles
in center-of-mass coordinates using three translational, three
rotational, $3N-6$ vibrational, and $3N$ internal degrees of freedom. As
in particle coordinates, there are a total of $6N$ positional degrees of
freedom and $6N$ associated momenta. However, each positional degree of
freedom now describes a collective motion of all the particles in the
cluster.

We define our coordinate system as follows. There is a lab frame with a
fixed origin and a rotating frame with an origin $O$ at the center of
mass of our cluster (Fig.~\ref{fig:CoordSystems}, right). We choose the
rotating frame such that the axes lie along the principal axes of the
cluster. For symmetric clusters, there may be more than one choice of
principal axes; we arbitrarily pick one set.

Six degrees of freedom describe the position and orientation of the
rotating frame relative to the lab frame. The translational degrees of
freedom $\mathbf{q}' = (q_x', q_y', q_z')$ describe the position of
$O$ relative to the lab origin. Their conjugate linear momenta are
$\mathbf{p}'=(p_x', p_y', p_z')$. The rotational degrees of freedom
$\bm{\Phi}'=(\phi', \theta', \psi')$ are the standard Euler angles.
Their conjugate angular momenta are $\mathbb{L}'= (p_\phi', p_\theta',
p_\psi')$.

With the harmonic approximation, we can describe the vibrations of the
cluster using a set of $3N-6$ orthogonal harmonic modes. The
displacement along the $j$th mode is $\xi'_j$, and the set of all
vibrational displacements is $\bm{\xi}'=(\xi'_1, \ldots, \xi'_{3N-6})$.
When all the particles are in their lowest potential-energy
configurations, $\bm{\xi}'=0$. The momentum conjugate to the vibrational
coordinate for the $j$th mode is $\chi'_j$, and the set of all
vibrational momenta is $\bm{\chi}'$.

Under these assumptions the Hamiltonian in center-of-mass coordinates
becomes
\begin{equation}\label{COM-Hamiltonian}
	\begin{split}
		\mathcal{H}_{s}'  & =  \mathcal{H}_{s, \textnormal{trans}}'+ 
    \mathcal{H}_{s, \textnormal{rot}}' + \mathcal{H}_{s, \textnormal{vib}}'
    + \mathcal{H}_\textnormal{rot} \\
    & = U_s(\mathbf{q}') + K_s(\mathbf{p}') + U_s(\bm{\Phi}') + K_s(\mathbb{L}')  \\
    & \quad + U_s(\bm{\xi}') + K_s(\bm{\chi}') + \mathcal{H}_\textnormal{rot} (\bm{\Phi}, \mathbb{L}),
  \end{split}
\end{equation}
where, for completeness, we have included a term describing the
rotations of individual particles. The partition function $Q_s'$ is then
\begin{equation}\label{COM-PartitionFunction}
	\begin{split}
    Q_{s}' = \frac{1}{\sigma_sh^{6N}} &\int {e^{-\beta \mathcal{H}_{s}'}} d \mathbf{q}' \, d\mathbf{p}' \, d\bm{\Phi}' \, d\mathbb{L}' \,
    d\bm{\xi}' \, d\bm{\chi}' \, d \bm{\Phi} \, d \mathbb{L} \\
          = \frac{Q_\textnormal{rot}}{\sigma_sh^{6N}} 
            &\int {e^{-\beta \mathcal{H}'_\textnormal{trans}}} \, d\mathbf{q}' \, d\mathbf{p}' 
            \int {e^{-\beta \mathcal{H}'_\textnormal{rot}}} \, d\bm{\Phi}' \, d\mathbb{L}'\\
            &\times\int {e^{-\beta \mathcal{H}'_\textnormal{vib}}} \, d\bm{\xi}' \, d\bm{\chi}' \\
        = \frac{Q_\textnormal{rot}}{\sigma_sh^{6N}} 
        & Q_{s, \textnormal{trans}}'(\mathbf{q}', \mathbf{p}') Q_{s, \textnormal{rot}}'(\bm{\Phi}', \mathbb{L}') Q_{s, \textnormal{vib}}'(\bm{\xi}', \bm{\chi}'),
	\end{split}
\end{equation}
where the last line defines the translational ($Q_{s,
  \textnormal{trans}}'$), rotational ($Q_{s, \textnormal{rot}}'$), and
vibrational ($Q_{s, \textnormal{vib}}'$) partition functions in
center-of-mass coordinates. As we did in particle coordinates, we
express the contribution of individual particle rotations as
$Q_{\textnormal{rot}}$, which is a constant for all structures formed
from the same $N$ particles. We also divide by the symmetry number
$\sigma_s$ to avoid overcounting rotational states. We discuss the role
of the symmetry number in more detail in
Section~\ref{sec:SymmetryNumber}.

The expression for the partition function in
Eq.~(\ref{COM-PartitionFunction}) is more tractable than the one in
particle coordinates [Eq.~(\ref{Particle-CompletePartitionFunction})]
because we need not set any cutoff distances that depend on the
structure. Instead, we can integrate over all possible values of the
translational, rotational, and vibrational coordinates.

The absence of bounds that depend on the structure raises the question
of where exactly we specify the structure when we calculate the
integral. The structure is in fact encoded in how the coordinates couple
to the energies. For example, different structures have different
harmonic modes, and these modes couple to the vibrational energy through
a set of natural frequencies that are different for each structure.
Furthermore, although the rotational modes of all the clusters are the
same---they represent rotations about three orthogonal axes---these
modes couple to the angular kinetic energy through the principal moments
of inertia, which differ from structure to structure. Below, we
analytically integrate the translational, rotational, and vibrational
partition functions and point out the terms that define the structure.

%%%%%%%%%%%%%%%%%
\subsubsection{Center-of-mass translations and linear momenta}
We start by calculating the translational partition function in
Eq.~(\ref{COM-PartitionFunction}), which is given by
\begin{equation*}
Q_{s, \textnormal{trans}}' = \int e^{-\beta U'_s(\mathbf{q}')} d\mathbf{q}' 
\int e^{-\beta K'_s(\mathbf{p}')} d\mathbf{p}'.
\end{equation*}
We assume that the potential energy of the cluster does not vary with
its position in space, such that $U'(\mathbf{q}')=0$.

The first integral in the translational partition function is then equal
to the volume available to the cluster, which is the total volume $V$ of
the container less any volume that the cluster cannot access without
penetrating a boundary. We can neglect this excluded volume if $V \gg
V_s$, where $V_s$ is some measure of the volume of a particular
structure. With this approximation,
\begin{equation*}
  \int e^{-\beta U'(\mathbf{q}')} d\mathbf{q}' = V.
\end{equation*}

The second integral in the translational partition function can also be
analytically integrated. The translational kinetic energy of the center
of mass of the cluster is given by
\begin{equation*}
	K'(\mathbf{p}') = \frac{1}{2M} (p_x'^2 + p_y'^2 + p_z'^2),
\end{equation*}
where $M=\sum_{i=1}^{N}{m_i}$ is the total mass of the cluster. After
evaluating the resulting Gaussian integral, we find that
\begin{equation*}
	\begin{split}
		Q_{\textnormal{s, trans}}' & = V \left( {\frac{2 \pi M}{\beta}} \right)^{3/2}.
    \end{split}
\end{equation*}

%%%%%%%%%%%%%%%
\subsubsection{Center-of-mass rotations and angular momenta}
\label{COM_Rotations}

The rotational partition function in center-of-mass coordinates
describes the free rotation of the cluster about its center of mass:
\begin{equation} \label{COM-InitialRotPartitionFunction}
Q_{s, \textnormal{rot}}' = \int e^{-\beta U'_s(\bm{\Phi}')} d\bm{\Phi}' 
\int e^{-\beta K'_s(\mathbb{L}')} d\mathbb{L}'.
\end{equation}  We assume that the potential energy of the cluster is
independent of its orientation, such that $U'(\bm{\Phi}') = 0$.  We need
not include a Jacobian term when the rotational partition function is
written in the form above, which is an integral over the the Euler
angles $\bm{\Phi}' = (\phi', \theta', \psi')$ and their conjugate
angular momenta $\mathbb{L}'$.

However, it is more natural to express the second integral in
Eq.~(\ref{COM-InitialRotPartitionFunction}) in terms of the angular
velocities $\bm{\Omega}'$ of the cluster about its principal axes:
\begin{equation*}
\begin{split}
&\Omega_1' = \dot{\theta}' \sin{\psi'} - \dot{\phi}'\sin{\theta'}\cos{\psi'}\\
&\Omega_2' = \dot{\theta}' \cos{\psi'} + \dot{\phi'}\sin{\theta'}\sin{\psi'}\\
&\Omega_3' = \dot{\psi}'+\dot{\phi'}\cos{\theta'},
\end{split}
\end{equation*}
where the dots denote time derivatives, and the subscripts denote
principal axes. The angular kinetic energy of the cluster is then
\begin{equation}\label{COM-Kinetic}
	K'_s(\bm{\Omega}') = \frac{1}{2} \left( I_{s, 1}\Omega_1'^2 + I_{s, 2}\Omega_2'^2 + I_{s, 3}\Omega_3'^2 \right),
\end{equation}
where $I_{s, 1}$, $I_{s, 2}$, and $I_{s, 3}$ are the principal moments
of inertia, which depend on the specific structure $s$.

Following A. Wilson~\cite{AWilson_1957}, we change variables from the
conjugate momenta $(p_\theta', p_\phi', p_\psi')$ to the angular
velocities $(\Omega_1', \Omega_2', \Omega_3')$. We calculate the
momenta from derivatives of the Lagrangian $\mathcal{L}'=K' - U'$ with
$U' = 0$ as discussed above:
\begin{equation}\label{COM-AngMomenta}
\begin{split}
  p_\theta' = \frac{\partial{K_s'}}{\partial{\dot{\theta}'}} &= I_{s, 1} \Omega_1' \sin{\psi'} + I_{s, 2} \Omega_2' \cos{\psi'}\\
  p_\phi' = \frac{\partial{K_s'}}{\partial{\dot{\phi}'}} &= -I_{s, 1} \Omega_1 '\sin{\theta'} \cos{\psi'} \\
  &\quad +I_{s, 2} \Omega_2' \sin{\theta'} \sin{\psi'} + I_{s, 3} \Omega_3' \cos{\theta'}\\
  p_\psi' = \frac{\partial{K_s'}}{\partial{\dot{\psi}'}} &= I_{s, 3}
  \Omega_3'.
\end{split}	
\end{equation}
The change of variables introduces a Jacobian term
\begin{equation*}
  J = \frac{\partial\left(p_\theta', p_\phi', p_\psi' \right) }{\partial \left(\Omega_1', \Omega_2', \Omega_3' \right) } = I_{s, 1} I_{s, 2} I_{s, 3} \sin{\theta'}.
\end{equation*}

We can then integrate to yield the rotational partition function:
\begin{equation}\label{COM-Rot Partition Function}
	\begin{split}
		Q_{s, \textnormal{rot}}' = &\int e^{-\beta U_s(\bm{\Phi}')} d\bm{\Phi}'  \int e^{-\beta K_s(\mathbb{L}')} d\mathbb{L}'\\
        = &\iint J e^{-\beta U_s(\bm{\Phi}')} e^{-\beta K_s(\bm{\Omega}')} d\bm{\Phi}' \, d\bm{\Omega}'\\
        = &\,I_{s, 1} I_{s, 2} I_{s, 3} \int \sin{\theta'}\, d\bm{\Phi}'\\
        &\times\int{ \exp{\left[ -\frac{\beta}{2} \left( I_{s, 1}\Omega_1'^2 + I_{s, 2}\Omega_2'^2 + I_{s, 3}\Omega_3'^2 \right)\right]}}  
        d \bm{\Omega}' \\
        = &\,8 \pi^2 \left( \frac{2\pi}{\beta} \right)^{3/2} \sqrt{I_{s, 1} I_{s, 2} I_{s, 3}},
    \end{split}
  \end{equation}
%%%%%%%%%%%%%%
\subsubsection{Vibrational modes}

Finally, we calculate the partition function associated with the
remaining $3N-6$ degrees of freedom. The vibrational partition function
is given by
\begin{equation*}
Q_{s,\textnormal{vib}}' = \int e^{-\beta U'_s(\bm{\xi}')} d\bm{\xi}' 
\int e^{-\beta K'_s(\bm{\chi}')} d\bm{\chi}'.
\end{equation*}

The harmonic modes that we use to describe the vibrations are the
eigenvectors of the mass-weighted Hessian (where the $ik$th entry is
scaled by $1/\sqrt{m_i m_k}$). We select only the $3N-6$ modes that have
non-zero eigenvalues. The $j$th eigenvector has an associated eigenvalue
that we denote $\omega_j$. Thus, the vibrational potential energy can be
expressed as a product of squared displacements along the modes:
\begin{equation*}
	\begin{split}
    	U_{s}'(\bm{\xi}') = U_0 + \sum_{j=1}^{3N-6}{\frac{\omega^2_{s,j}}{2} \xi_j'^2},
    \end{split}
\end{equation*}
where $\xi_j'$ is the displacement along the $j$th mode and $U_0$ is the
total potential energy in the absence of vibrational excitations. We set
$U_0=0$ from here on.

The vibrational kinetic energy is
\begin{equation*}
  K_{s}'(\bm{\chi}') = \sum_{j=1}^{3N-6}\frac{1}{2}\chi_j'^2,
\end{equation*}
where $\chi_j$ is the (mass-weighted) momentum along the $j$th mode.

We can then analytically integrate the vibrational partition function to
obtain
\begin{equation*}
	\begin{split}
		Q_{s, \textnormal{vib}}'  = &\int{ \exp{\left(-\frac{\beta}{2} 
        \sum_{j=1}^{3N-6} {\omega^2_{s,j} \xi_j'^2} \right) }}d\bm{\xi}' \\
         &\times\int \exp \left( -\frac{\beta}{2} \sum_{j=1}^{3N-6} \chi_j'^2 \right) d\bm{\chi}' \\
        =  &\prod_{j=1}^{3N-6} \frac{2\pi}{\beta \omega_{s,j}}.
    \end{split}
\end{equation*}

%%%%%%%%%%%%%%%%%%%%%%%%%%%%%%
\subsubsection{Complete partition function in center-of-mass coordinates}

Putting together the translation, rotational, and vibrational components
with the prefactor in Eq.~(\ref{COM-PartitionFunction}), we obtain the
complete partition function of a structure $s$ in center-of-mass
coordinates:
\begin{widetext}
  \begin{equation}\label{COM-CompletePartitionFunction}
    	Q_s' = \frac{Q_\textnormal{rot}}{\sigma_s h^{6N}} { 8 \pi^2 V M^{3/2}}
      \left( {\frac{2\pi}{\beta}} \right)^{3N-3} \sqrt{I_{s, 1} I_{s, 2}
        I_{s, 3}} \left( \prod_{j=1}^{3N-6} \frac{1}{\omega_j} \right)
  \end{equation}
\end{widetext}
The version for quasi-two-dimensional systems is given in the Appendix.
We note that $Q_\textnormal{rot}$, which accounts for the rotations of
individual particles, will cancel in the calculation of the cluster
probabilities. Apart from the factor of $Q_\textnormal{rot}/h^{3N}$,
Eq.~(\ref{COM-CompletePartitionFunction}) is equivalent to the molecular
partition function, in that the same expression can be derived from the
quantum version of the Hamiltonian by taking the high-temperature limit.
In this limit, no modes are frozen out.

It is much easier to calculate an explicit value of the partition
function with the center-of-mass formulation,
Equation~(\ref{COM-CompletePartitionFunction}), than with the particle
coordinate formulation, Eq.~(\ref{Particle-CompletePartitionFunction}).
Apart from $Q_\textnormal{rot}$, all the constants in
Eq.~(\ref{COM-CompletePartitionFunction})---the moments of inertia, the
vibrational frequencies, and the total mass---can be calculated directly
from the positions and masses of the particles. By contrast, in
Eq.~(\ref{Particle-CompletePartitionFunction}), we must calculate the
volumes associated with all fluctuations of the structure. Calculating
these effective volumes requires calculating a Jacobian for each
particle~\cite{herschbach59, boresch96}, because the momenta are already
integrated out. By using the rigid-rotor-harmonic-oscillator
approximation and taking advantage of the separability of translations,
rotations, and vibrations, we are largely able to avoid the calculations
of Jacobians in Eq.~(\ref{COM-CompletePartitionFunction}). In the
vibrational partition function, for example, we use a coordinate system
natural to the vibrational modes (and different from that for the
rotational modes) and pair the positions along the modes with their
conjugate momenta.

However, this separability comes at a cost. Certainly it sacrifices
generality: Equation~(\ref{Particle-CompletePartitionFunction}) does not
rely on the rigid-rotor-harmonic-oscillator approximation, whereas
Eq.~(\ref{COM-CompletePartitionFunction}) does. The results of
experiments do agree with the predictions of
Eq.~(\ref{COM-CompletePartitionFunction}), establishing the validity of
the approximation. But the more serious problem with
Eq.~(\ref{COM-CompletePartitionFunction}) is that it obscures the
essential physics. It is written in terms of moments of inertia and
vibrational frequencies---dynamical parameters whose names suggest that
inertia and vibrations can affect the equilibrium probability. As we
discuss below, the terminology associated with these quantities can lead
to confusion.

%%%%%%%%%%%%%%%%%%
\section{Discussion}
\label{sec:discussion}

%%%%%%%%%%%%%%%%%%%%%%%%
\subsection{Moments of inertia and vibrational frequencies}\label{MassDependenceDerivation}

\begin{figure}
	\begin{centering}
    \includegraphics{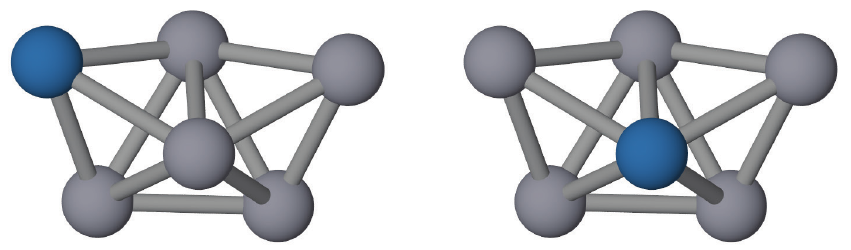}
    \caption{Schematic of a tri-tetrahedral cluster with one particle
      (blue) that is heavier than the others, but otherwise identical.
      In this rendering, the pair interactions, or bonds, are shown as
      struts connecting the particles, which are not drawn to scale.
      Although the moment of inertia decreases if the particle moves
      from the position shown on the left to that shown on the right,
      the equilibrium probability of the cluster does not change, as
      discussed in the text.}
    \label{fig:Mass_Difference}
	\end{centering}
\end{figure}

Equation~(\ref{COM-CompletePartitionFunction}) might seem to suggest
that the value of the partition function would differ if we switched the
location of a massive particle with that of a lighter one in the same
structure. Say we have a structure $s$ and a set of particles, all of
which have the same sizes and interactions, but one of which is much
denser than the others. A cluster with the denser particle located near
the center of mass will have much lower moments of inertia than a
cluster with the particle located further away, as shown in
Fig.~\ref{fig:Mass_Difference}. Therefore the value of the rotational
partition function [Eq.~(\ref{COM-Rot Partition Function})] will be much
smaller when the particle is closer to the center of mass. We might then
expect that in equilibrium, such a configuration would occur less often
than a configuration with the particle further from the center of mass.
This behavior is counterintuitive, because we expect inertia to have no
effect in a typical colloidal suspension, where the surrounding liquid
damps the motion~\cite{cates12}.

This apparent dependence on the location of the masses is an artifact of
the separation of the Hamiltonian into rotational and vibrational
components. In particle coordinates, where we do not separate the
Hamiltonian, the value of the partition function clearly does not depend
on the location of the particles:
Equation~(\ref{Particle-CompletePartitionFunction}) shows that only the
product of the masses matters. Thus, any changes to the moments of
inertia resulting from switching the masses must be compensated by
changes in the vibrational frequencies.

We can demonstrate this invariance to the positions of the masses by
equating the partition function in center-of-mass coordinates,
Eq.~(\ref{COM-CompletePartitionFunction}), to that in particle
coordinates, Eq.~(\ref{Particle-CompletePartitionFunction}). The value
of the partition function for a given structure $s$ should be the same
in both coordinate systems if the rigid-rotor-harmonic-oscillator
approximation is valid. Several terms cancel when we equate the two,
including $Q_\textnormal{rot}$, $\sigma_s$, and $h$.

The remaining terms can be sorted into two groups: those that depend on
the particle masses, and those that do not. The terms that depend on the
masses are the sum of masses $M$, the product of the moments of
inertia $I_{s, 1} I_{s, 2} I_{s, 3}$, and the product of the vibration
frequencies $\prod_{j=1}^{3N-6}\omega_j$. Terms that do not depend on
the masses include the volume $V$, the inverse thermal energy $\beta$,
and the configurational partition function $Z_s$, which, for a
non-gravitational potential, depends only on the interactions and the
positions of the particles and not their masses. We can further group
the terms that depend implicitly on the masses---the moments of inertia
and the vibrational frequencies---on one side of the equation, and the
terms that depend explicitly on the masses---$M$ and $\prod_{i=1}^N
m_i$---on the other~\cite{perry_thesis}. We then find:
\begin{equation}\label{RotationalVibrationalProduct}
	\begin{split}
    	&f(V, \beta, Z_s) \sqrt{I_{s, 1} I_{s, 2} I_{s, 3}} 
        \left( \prod_{j=1}^{3N-6} \frac{1}{\omega_j} \right) = \frac{\prod_{i=1}^{N} m_i^{3/2}}{M^{3/2}}
    \end{split}
\end{equation}
where $f(V, \beta, Z_s)$ is a function that depends neither implicitly
nor explicitly on the masses. The form of
Eq.~(\ref{RotationalVibrationalProduct}) agrees with that derived by
Herschbach, Johnston, and Rapp \cite{herschbach59}.

We have therefore shown that the product of the moments of inertia and
the inverse vibrational frequencies is proportional to the ratio of a
product and a sum of the masses:
\begin{equation}
    \sqrt{I_{s, 1} I_{s, 2} I_{s, 3}} \left( \prod_{j=1}^{3N-6} \frac{1}{\omega_j} \right) \propto
    \frac{\prod_{i=1}^{N} m_i^{3/2}}{\left( \sum_{i=1}^{N} m_i \right)^{3/2}},
\end{equation}
Because the products and sums on the right side are invariant to
permutations, the product on the left side must also be invariant to the
positions of the masses, so long as the structure remains the same.

The discussion above shows that we should consider the moments of
inertia as geometrical or structural quantities rather than dynamical
ones, at least for the purposes of calculating the partition function. A
moment of inertia characterizes the geometrical extent of a
cluster---the larger the moment, the larger the radius of gyration of
the cluster, and the larger the effective volumes it would sweep out in
particle coordinates [Eq.~(\ref{Particle-CompletePartitionFunction})].
Larger moments of inertia correspond to higher entropy.

We also interpret the vibrational frequencies as structural rather than
dynamical quantities. Their appearance in the partition function does
not mean that the particles actually oscillate. The $\omega_j$ appear as
a shorthand for the non-zero eigenvalues of the mass-weighted Hessian,
and, as such, account for how the structure determines the vibrational
potential energy.

%%%%%%%%%%%%%%%%%%
\subsection{Symmetry}\label{sec:SymmetryNumber}

As discussed in Section~\ref{sec:background}, structures with low
symmetry are favored in equilibrium when $N<9$. This result is exactly
as predicted by the statistical mechanical model above: the partition
function in either particle or center-of-mass coordinates is inversely
proportional to the symmetry number. As a consequence, we expect that in
an equilibrium ensemble, structures with lower symmetry occur more often
than those with higher symmetry.

To understand why the symmetry number $\sigma_s$ appears in the
partition function, we must consider how experimentalists measure the
equilibrium probabilities. Meng and coworkers~\cite{meng10} made an
equilibrium ensemble of clusters and, using a microscope, took videos of
each cluster as it rotated and translated owing to Brownian motion. They
identified the structure of each cluster from the videos by visual
inspection in the case of the octahedron and tri-tetrahedron, or by
determining the networks of contacts and the adjacency matrix in the
case of more complicated structures. In either case, they did not
distinguish the particles. Finally, to obtain the equilibrium
probabilities, they counted the number of times each different structure
appeared in the ensemble for a given number of particles.

For a statistical mechanical model to reproduce the experimentally
measured probabilities, it must ``count'' clusters in a similar
way---independently of their orientation. An example of a model that
does \emph{not} fit this criterion is one in which we define the bounds
on the partition function to include only one particular orientation of
a structure. For example, we might include only the orientation of the
octahedron with a triangular face facing toward us and a vertex of that
face pointing down, as shown in Fig.~\ref{fig:Octahedron_Colorings}.
There are 24 ways in which an octahedral cluster can attain this
orientation, owing to its symmetry. We illustrate the 24 different ways
by giving different colors to the particles in
Fig.~\ref{fig:Octahedron_Colorings}. By contrast, the same accounting
for a tri-tetrahedron would show that there are only two ways of
obtaining the same orientation. Thus, our faulty partition function
would overcount the octahedron by a factor of $12=24/2$.

To correct our faulty model, we might integrate over all orientations.
But in doing so, we must correct for the overcounting of states at any
particular orientation. Put another way: the rotational partition
function in center-of-mass coordinates, Eq.~(\ref{COM-Rot Partition
  Function}), extends over all Euler angles. But for a given set of
principal axes, there are 24 equivalent choices of Euler angles for any
orientation of the octahedral cluster. This factor of 24 is the symmetry
number $\sigma_s$ that we include in the denominator of the partition
function, as shown in Eq.~(\ref{COM-CompletePartitionFunction}). As a
result, our corrected model predicts that the equilibrium probability of
the octahedron relative to the tri-tetrahedron is proportional to
$\sigma_\textnormal{tri-tetrahedron}/\sigma_\textnormal{octahedron}$.

\begin{figure}
	\begin{centering}
    \includegraphics{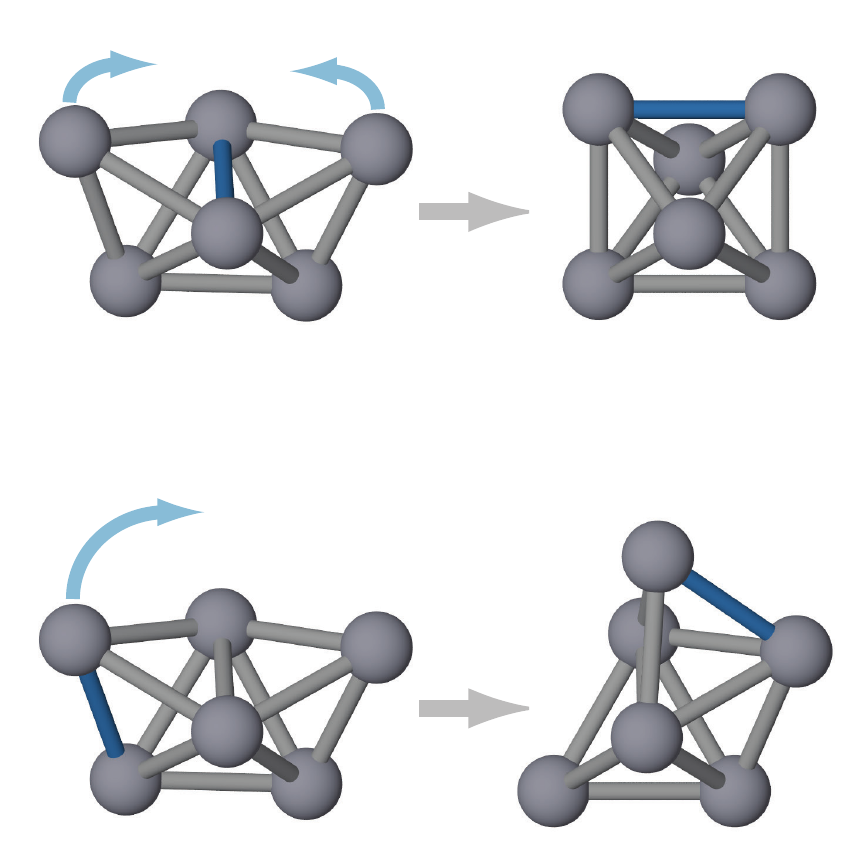}
    \caption{Detailed balance argument for how symmetry affects
      probability.  Top: There is only one bond in the tri-tetrahedron (blue
      bond in top left) that, once broken, allows the structure to
      transition to the octahedron. By contrast, breaking any of the 12
      symmetrically-equivalent bonds in the octahedron allows it to
      transition back to the tri-tetrahedron. Bottom: Breaking any other
      bond in the tri-tetrahedron, such as the blue bond shown, leads
      back to the tri-tetrahedron.
      \label{fig:Detailed_Balance}}
	\end{centering}
\end{figure}

The argument above explains the mathematical reason for the symmetry
number, but it does not explain its apparent \emph{physical}
effect---suppressing the occurrence of highly symmetric clusters like
the octahedron. This effect is most easily explained using detailed
balance. Let us neglect any fluctuations in bond distances and consider
only the ways in which bonds can break. There are 12 symmetrically
equivalent bonds in the octahedron, and breaking any of them allows the
octahedron to transition to the tri-tetrahedron. By contrast, there is
only one bond that, once broken, allows the tri-tetrahedron to
transition back to the octahedron, as shown in
Fig.~\ref{fig:Detailed_Balance}. Detailed balance then requires that the
equilibrium probability of the tri-tetrahedron be a factor of 12 higher
than that of the octahedron. This factor is exactly the ratio of
symmetry numbers.

To account for the factor of 24 measured in the experiments, we must
include contributions from the rotational and vibrational partition
functions as well. But the above arguments show clearly that the
symmetry number accounts for an entropic effect: there are more ways to
arrange particles into a low-symmetry structure like a tri-tetrahedron
than a high-symmetry one like the octahedron.

The symmetry number therefore does not account for whether the particles
are fundamentally distinguishable or not. Nor does its origin lie in
quantum mechanics. It arises because different orientations of a cluster
are not counted as different states in the experiment.

%%%%%%%%%%%%%%%%%%%%%%
\section{Conclusions and more questions}

We have shown that for isostatic clusters composed of undistinguished
particles with a harmonic potential, the partition function for
colloidal systems is equivalent to the high-temperature limit of the
molecular partition function. We have also shown that the effects of all
of its terms on the equilibrium cluster probabilities can be explained
classically. By equating the partition functions in particle and
center-of-mass coordinates, we have shown that the moments of inertia
and vibrational frequencies should be interpreted as geometrical or
structural quantities rather than dynamical ones, at least for the
purposes of calculating the equilibrium probabilities. We have
underscored this point by showing that the ostensible dependence on the
positions of the masses in center-of-mass coordinates is an artifact of
the separation between rotational and vibrational modes.

The model we derive can be applied to other systems if we relax some of
our assumptions. For example, it can be applied to clusters of
anisotropic particles~\cite{hong_clusters_2008, chen11a} in which one
\emph{does} distinguish clusters by the orientations of particles within
them. In such systems, there is a rotational potential energy term that
causes particles to favor certain orientations over others. Thus the
$Q_\textnormal{rot}$ term will depend on the structure and will not
cancel in the observation probabilities.

If we relax the rigid-rotor-harmonic-oscillator approximation, we can
begin to describe even more classical systems. For all non-relativistic
classical systems with a non-gravitational potential, the form of the
partition function in particle coordinates remains the same as what we
have derived---so long as it makes sense to define the structure in
terms of an adjacency matrix. Thus, for all non-gravitational,
non-relativistic classical systems where this structural description is
valid, the masses and the positions decouple in the partition function.

Finally, we note that although we have focused on isostatic, rigid
colloidal clusters, the singular and hyperstatic clusters are important
to understand because they can occur with high probabilities in
experiments \cite{meng10}---singular clusters because of their high
vibrational entropy and hyperstatic clusters because of their low
potential energy. The center-of-mass partition function we derive here
breaks down in these cases. However, the form of the partition function
in particle coordinates is still valid. Kallus and Holmes-Cerfon have
developed a theoretical framework to calculate the equilibrium
probabilities of such structures~\cite{kallus_free_2017}.

The overriding goal of both statistical mechanical models and
experiments on colloidal clusters of spheres is to understand how the
free-energy landscape changes as a function of the number of particles
$N$. The landscape quantifies the frustration of the system and how that
frustration evolves in the limit as $N\to\infty$, where we expect that
the ground state is a crystal. To this end, our model and the physical
interpretations we give are important because they give insights into
the depths of the minima on the landscape. Although the vibrational
framework we use breaks down for non-rigid clusters, the invariance to
the positions of masses as well as the entropic effects of the moments
of inertia and symmetry number are valid for all clusters. Understanding
their physical effects is crucial to making sense of the complex
landscape that emerges as $N$ increases.

%%%%%%%%%%%%%%%%%%%%%%%%
\appendix*
\section{Partition function in two dimensions}

Here we show the result for the partition function for a two-dimensional
(2D) cluster, which differs from the three-dimensional (3D) result
because the degrees of freedom are different in two dimensions. While
true 2D colloidal systems do not exist, we can model experiments in
which 3D spherical particles are confined to a 2D surface. For example,
under depletion interactions, planar clusters of spherical particles can
form at a surface such as a microscope
coverglass~\cite{perry_two-dimensional_2015}. As in three dimensions,
the depletion interaction does not prevent the particles in such a
quasi-2D cluster from rotating about their centers of mass. However, the
collective motions of the cluster are constrained to the plane defined
by the surface. Our 2D partition function is specific to such systems.

In particle coordinates, each particle has two positional degrees of
freedom. The partition function for a structure $s$ confined to a plane
is given by
\begin{equation}\label{2DParticle-CompletePartitionFunction}
	\begin{split}
		Q_{s}^{2D} 
        & = \frac{Q_\textnormal{rot}}{\sigma_s^{2D} h^{5N}} \left( \frac{2 \pi}{\beta} \right)^{N}  Z_{s}^{2D} \prod_{i=1}^{N} m_i \\
      & = \frac{Q_\textnormal{rot}}{\sigma_s^{2D} h^{5N}} \left( \frac{2 \pi}{\beta} \right)^{N}  \left( A\prod_{i=2}^N A_i \right)\prod_{i=1}^{N} m_i,
   \end{split}
\end{equation}
where in the second line we have replaced $Z_{s}^{2D}$ by a product of
areas for each particle, analogous to the product of volumes in
Eq.~(\ref{Particle-CompletePartitionFunction}). The factor of $Q_{rot}$
is the same as that in the 3D case because our particles are still free
to rotate about their own centers of mass in all three dimensions. The
particle rotations also contribute a factor of $1/h^{3N}$ to the
partition function, with the remaining factor of $1/h^{2N}$ coming from
the translations of individual particles. The symmetry number
$\sigma_s^{2D}$ differs from that in three dimensions because it does
not account for out-of-plane rotations.

In center-of-mass coordinates, the colloidal cluster has two
translational degrees of freedom and only one rotational degree of
freedom, since the only allowed rotations are in the plane. An isostatic
cluster has $2N-3$ vibrational modes. The center-of-mass partition
function for a 2D cluster is then
\begin{equation}\label{2DCOM-CompletePartitionFunction}
  Q_{s}^{\prime 2D} = \frac{Q_\textnormal{rot}}{ \sigma_{s}^{2D}h^{5N}} 2\pi A M \sqrt{I_{s}}
  \left( {\frac{2\pi}{\beta}} \right)^{(4N-3)/2}  \left( \prod_{j=1}^{2N-3} \frac{1}{\omega_j} \right).
\end{equation}

\begin{acknowledgments}
  We thank Miranda Holmes-Cerfon, Michael Brenner, Manhee Lee, Carl
  Goodrich, Abigail Plummer, Sarah Kostinski, Mike Cates, and Tom Witten
  for helpful discussions. Rebecca W. Perry and Ellen D. Klein
  acknowledge the support of National Science Foundation (NSF) Graduate
  Research Fellowships. This work was funded by the NSF through grant
  no. DMR-1306410.
\end{acknowledgments}

\bibliography{manuscript}

\end{document}